\documentclass{raa_twocolumn} 
\usepackage{graphicx,times}             %for PS/EPS graphics inclusion, new
\usepackage{natbib}
\usepackage{amssymb,amsmath}
\usepackage{longtable}
\usepackage{subcaption}
\usepackage{graphicx}
\usepackage{multirow}
\usepackage{cancel}
\usepackage{pdflscape}
\usepackage{grffile}
\usepackage{float}
\usepackage{xcolor} 
\bibpunct{(}{)}{;}{a}{}{,}
\usepackage[a4paper=true,pagebackref=true]{hyperref}
\hypersetup{colorlinks = true, linkcolor = blue, anchorcolor = blue, citecolor = blue, filecolor = blue, pagecolor = blue, urlcolor = blue}

\usepackage{CJK}
\usepackage{upgreek}

\usepackage{threeparttable, longtable, threeparttablex, adjustbox, booktabs, url, pdflscape}

\def\be{\begin{equation}}
\def\ee{\end{equation}}

\begin{document}

\title{On the energy budget of starquake-induced repeating fast radio bursts}
\volnopage{ {\bf 20xx} Vol.\ {\bf xx} No. {\bf x}, XXX}
\setcounter{page}{1}

\begin{CJK*}{UTF8}{gbsn}
\author{
    Wei-Yang Wang (王维扬)\inst{1,2}, Chen Zhang (张晨)\inst{3}, Enping Zhou (周恩平)\inst{4}, Xiaohui Liu (刘小辉)\inst{5,1}, 
    Jiarui Niu (牛佳瑞)\inst{5,1}, Zixuan Zhou (周子轩)\inst{2,6},
    He Gao (高鹤)\inst{7}, Jifeng Liu (刘继峰)\inst{1,8,9}, Renxin Xu (徐仁新)\inst{2}, Bing Zhang (张冰)\inst{10,11}
}
\institute{
    School of Astronomy and Space Science, University of Chinese Academy of Sciences, Beijing 100049, China; {\it wywang@ucas.ac.cn} \\
    \and
  State Key Laboratory of Nuclear Physics and Technology, School of Physics, Peking University, Beijing 100871, China; {\it r.x.xu@pku.edu.cn}\\
  \and 
  The HKUST Jockey Club Institute for Advanced Study, The Hong Kong University of Science and Technology, Hong Kong, China; {\it iasczhang@ust.hk}\\
  \and
  School of Physics, Huazhong University of Science and Technology, Wuhan 430074, China\\
  \and
  National Astronomical Observatories, Chinese Academy of Sciences, Beijing 100101, China\\
  \and
  School of Mathematics and Physics, University of Science and Technology, Beijing 100083, China\\
  \and
  Department of Astronomy, Beijing Normal University, Beijing 100875, China\\
  \and
  New Cornerstone Science Laboratory, National Astronomical Observatories, Chinese Academy of Sciences, Beijing 100101, China\\
\and
Institute for Frontiers in Astronomy and Astrophysics, Beijing Normal University, Beijing 102206, China\\
\and
Nevada Center for Astrophysics, University of Nevada, Las Vegas NV 89154, USA\\ 
\and
Department of Physics and Astronomy, University of Nevada, Las Vegas NV 89154, USA\\
  \vs \no
  {\small Received 20xx XXX; accepted 20xx XXX}
}
\abstract{
With a growing sample of fast radio bursts (FRBs), we investigate the energy budget of different power sources within the framework of magnetar starquake triggering mechanism.
During a starquake, the energy can be released in any form through strain, magnetic, rotational, and gravitational energies.
The strain energy can be converted from other three kinds of energy during starquakes.
The following findings are revealed:
1. The crust can store free magnetic energy of $\sim10^{46}$ erg by existing toroidal fields, sustaining $10^6$ bursts with frequent starquakes occurring due to crustal instability.
2. The strain energy develops as a rigid object spins down, which can be released during a global starquake accompanied by a glitch. However, it takes a long time to accumulate enough strain energy via spindown.
3. The rotational energy of a magnetar with $P\lesssim0.1\rm\,s$ can match the energy and luminosity budget of FRBs.
4. The budget of the total gravitational energy is high, but the mechanism and efficiency of converting this energy to radiation deserve further exploration.
}
\authorrunning{W. -Y. Wang et al.}
\titlerunning{FRB: Energy release of Starquake}
\maketitle

\section{Introduction}
\label{sec:intro}
\end{CJK*}

Fast radio bursts (FRBs) are millisecond-duration flashes, signifying extremely coherent radiation due to their extremely high bright temperatures.
Although they are similar in some aspects to single pulses from radio pulsars, the energy budget of FRBs is much higher. As a result, the physical origin(s) of FRBs is still a mystery \citep{2023RvMP...95c5005Z}.

A strong connection between magnetars and FRBs was established since the discovery of FRB 20200428D, an FRB-like burst from a Galactic magnetar, SGR J1935+2154 \citep{2020Natur.587...59B,2020Natur.587...54C}.
Based on the distance from the emission site to the central engine, magnetar FRB models can be generally divided into two categories: emission within the magnetosphere (pulsar-like models, \citealt{2014PhRvD..89j3009K,2017MNRAS.468.2726K,2018ApJ...868...31Y,2019ApJ...876L..15W,2020MNRAS.498.1397L,2021MNRAS.508L..32C,2022ApJ...927..105W,2022ApJ...925...53Z,2023ApJ...958...35L,2024A&A...685A..87W,2024arXiv240411948Q}), and emission from a relativistic shock region far outside the magnetosphere (GRB-like models, \citealt{2019MNRAS.485.4091M,2020ApJ...896..142B,2020ApJ...899L..27M,2022arXiv221013506C,2022ApJ...927....2K}).
The observations of polarization measurements, including diverse polarization angle swings \citep{2020Natur.586..693L} and high degrees of circular polarization \citep{2022RAA....22l4003J,2022Natur.609..685X,2023ApJ...955..142Z}, favor the pulsar-like models \citep{2023MNRAS.522.2448Q}, even though there are also some attempts to interpret these phenomena within the GRB-like models \citep{2024PhRvL.132c5201I}.

Starquakes from magnetars have been proposed as a promising triggering mechanism for FRBs \citep{2018ApJ...852..140W,2019MNRAS.488.5887S,2019ApJ...879....4W,2020RAA....20...56J,2022MNRAS.517.4612L}.
The FRB burst sequences share remarkably similar characteristics in energy distribution and temporal occurrence with Earthquakes \citep{2023arXiv230504738D,2023MNRAS.526.2795T,2023ApJ...949L..33W,2024MNRAS.tmp..976T}.
The quake model predicts that FRBs are associated with glitches and quasi-periodic oscillations (QPOs).
Surprisingly, a giant glitch was measured days before FRB 20200428D \citep{2020ApJ...898L..29M,2021NatAs...5..378L,2021NatAs...5..372R}, and later another glitch was found to be accompanied by three FRB-like radio bursts in subsequent days \citep{2023NatAs...7..339Y}.
A QPO of 40 Hz was reported from an X-ray burst during the active epoch of FRB 20200428D \citep{2022ApJ...931...56L}.

Several FRB sources have been observed to repeat more than a thousand times \citep{2021Natur.598..267L,2022Natur.609..685X,2023ApJ...955..142Z}, suggesting that an active central engine and a large energy reservoir are needed. The prospect of detecting more than 10 thousand bursts from an FRB source would pose significant challenges on the potential energy budget of a magnetar.
In this paper, we investigate the energy budget and the mechanism that can be released, within the framework of the starquake model for a neutron star or a magnetar.
We consider the energy budget of repeating FRBs in Section \ref{sec2}.
We discuss the energy release during starquakes by invoking strain, magnetic, rotational, and gravitational energy in Section \ref{sec3}.
The results are discussed in Section \ref{sec4} and summarized in Section \ref{sec5}.

\section{Energy budget of an FRB source}
\label{sec2}

As a cosmological burst, an FRB can release a huge amount of energy.
For a nominal Gpc distance $D$, %corresponding, 
the burst energy of an individual burst can be estimated as
\be
\begin{aligned}
\delta E& \simeq 4\pi D^2f_b F_\nu \Delta \nu\\
&=1.2\times 10^{38}f_b \,{\rm erg}\left(\frac{D}{1\,\rm Gpc}\right)^2\left(\frac{F_\nu}{0.1\,\mathrm{Jy}\,\mathrm{ms}}\right) \left(\frac{\Delta \nu}{1\,\mathrm{GHz}}\right),
\end{aligned}
\label{eq:E_obs}
\ee
where $f_b=\delta\Omega/(4 \pi)$ is the beaming factor for an individual burst, $F_\nu$ is the received fluence and $\Delta\nu$ is the bandwidth of the telescope (for a narrow FRB spectrum).
The emission solid angle $\delta\Omega$ could be as small as $1/(\gamma^2)\sim (10^{-4}-10^{-6})$ due to relativistic beaming \citep{2022ApJ...927..105W,2023RvMP...95c5005Z}.
Even though the energy of individual burst is smaller by a factor of $f_b$ than the isotropic value, there is a factor of $f_b^{-1}$ more FRBs that do not beam toward Earth. So the total energy budget of an FRB remains the same as estimated based on the observations \citep{2021Natur.598..267L,2022Natur.609..685X,2022RAA....22l4001Z} because the two factors cancel out.

However, there might be a global emission beam factor.
The bursts may not be emitted isotropically.
The emission cone of individual bursts is confined to a global fan-beam which has a solid angle of $\Delta\Omega$\footnote{Conal regions from a magnetospheric rotator are considered as the global beaming scenario, but a more general geometry is possible.}, leading to a global beam factor of $F_b=\Delta\Omega/(4\pi)$, with $\Delta\Omega \gg \delta\Omega$.
In this case, there is only a small fraction of bursts detected with altogether $\Delta\Omega/\delta\Omega$ bursts most of that are missed, so that the global energy budget can be smaller by a factor of $F_b$ from the isotropic value \citep{2023RvMP...95c5005Z}.
Therefore, when considering the total burst energy/luminosity, the global beaming factor rather than the beaming factor of an individual burst should be adopted.

It is necessary to introduce the luminosity/energy function of FRBs when a large sample of FRBs is considered.
Some statistical studies show that the luminosity/energy function of bursts from the same source can be characterized by a power law distribution, $N(E)\propto E^{-p}$ with indices ranging from $1-2$ \citep{2021Natur.598..267L,2022Natur.609..685X,2022RAA....22l4001Z,2023ApJ...955..142Z}.
For the FRB population involving different sources, the energy distribution can also be described by power laws with the similar index range \citep{2018ApJ...858...89C,2018MNRAS.481.2320L,2019ApJ...883...40L,2020MNRAS.498.1397L,2020MNRAS.494..665L}.
If the index is not strictly equal to 1 or 2, the average burst energy is calculated to 
\begin{equation}
\begin{aligned}
\bar{E}&=\frac{\int^{E_h}_{E_l}N(E)EdE}{\int^{E_h}_{E_l}N(E)dE}\\
&\simeq\frac{p-1}{p-2}\times\left\{\begin{array}{l}
E_h,~p<1\\
-E_h^{2-p} E_l^{p-1},~1<p<2\\
E_l, ~p>2
\end{array}\right..   
\end{aligned}
\label{eq:Ebar}
\end{equation}
The average burst energy mainly depends on the high-luminosity events when $p>2$ while low-luminosity events for $p<1$.
A central value of 1.8 can cover at least 7 orders of magnitude of burst energy \citep{2020MNRAS.498.1397L}.
For this power law index, the average energy or fluence for a repeater mainly depends on the low-energy events.
We take the observed fluence extending from $10^{-2}$ Jy ms to 100 Jy ms.
The average fluence for a repeater is calculated to 0.25 Jy ms based on Equation (\ref{eq:Ebar}).

The energy release may mainly support radiation in the X-ray bands.
Consider the peculiar event FRB 20200428D, the X-ray burst was more than $10^4$ more energetic than the associated radio burst \citep{2021Natur.598..267L,2020ApJ...898L..29M,2021NatAs...5..372R}, leading to an upper-limit radio efficiency of $\eta=10^{-4}-10^{-5}$ for this event.
Lower limits on $\eta$ by other X-ray counterparts are of this order or even smaller \citep{2021A&A...656L..15P}.
By multiplying the efficiency factor of $10^{-4}$, one finds that the theoretical total energy budget of a repeating FRB is at least $10^{46}\rm\, erg$.

Note that most bursts are missing due to the duty cycle.
The total observed number and their energy budget of some actively repeating FRBs within duty cycle are summarized in Table \ref{tab1}.
The physical duty cycle and burst rate evolution are complex.
We assume that the repeater source has a lifetime $\tau_{\rm life}$ and consider the average bursting rate to be independent of energy.
If the activity level remains unchanged during the lifetime, the total energy budget of the source over the lifetime can be estimated as \citep{2023RvMP...95c5005Z}
\be
E_{\mathrm{src}}=\int_0^{\tau_{\rm life}} \int_{E_{\text {iso }, \mathrm{m}}}^{E_{\mathrm{iso}, \mathrm{M}}}\left(\frac{d N}{d t d E_{\text {iso }}}\right) F_b E_{\mathrm{iso}} \eta^{-1} \zeta^{-1} d E_{\mathrm{iso}} d t,
\ee
where $\zeta=\tau_{\text {obs}} / \tau_{\rm life}$ is the observational duty cycle and $E_{\rm iso}$ is the isotropic FRB energy from the source.
A naive estimation is that considering the detection of 1652 bursts in $\sim60$ hours during a 47-day observational campaign \citep{2021Natur.598..267L}, leading to a $\zeta=0.053$.
The theoretical total energy budget of an observed FRB is required to exceed $\sim10^{43}$ erg at least and the total burst number may exceed $4\times10^4$.

\begin{table*}
\centering
\caption{Total observed number and total energy of some active repeaters}
\begin{tabular}{llll}
\hline \hline
FRB Name & Burst number & Total energy (erg) & Reference\\
\hline
20121102A & 2370 & $4.1\times10^{41}$ & \cite{2016Natur.531..202S,2018Natur.553..182M,2018ApJ...866..149Z};\\
& & & \cite{2021ApJ...908L..10H,2021Natur.598..267L,2022MNRAS.515.3577H}\\
20190520B & 121 & $1.1\times10^{40}$ & \cite{2022arXiv220308151D,2022AAS...24022601T,2022Natur.606..873N}\\
20201124A & 2883 & $1.6\times10^{41}$ & \cite{2022MNRAS.512.3400K,2022ApJ...927...59L,2022Natur.609..685X,2022RAA....22l4001Z}\\
20220912A & 1077 & $7.4\times10^{41}$ & \cite{2023ApJ...955..142Z}\\
\hline \hline
\end{tabular}
Some data are quoted from https://www.chime-frb.ca/repeaters/
\label{tab1}
\end{table*}

\section{Energy release during starquake}
\label{sec3}

\subsection{Strain energy from crustal cracking}\label{sec3.1}
Starquakes as a leading scenario to trigger FRBs can suddenly release a huge amount of energy.
It has been discussed as the energy source for some high-energy phenomena since the elastic energy released in the crust excites oscillations and the induced electric field accelerates charges to stream outflow \citep{1975A&A....44...21T,1976Ap&SS..42...77F,1986Ap&SS.120...27M,1988PhR...163..155E,1989ApJ...343..839B}.
The strain energy released in the crack region of $l^2$ is given by
\be
\begin{aligned}
\delta E_{\rm str}&=2\pi l^2\Delta R\mu\epsilon^2\\
&=4.7\times10^{37}\,{\rm erg}
\left(\frac{l}{10^4\,\rm cm}\right)^2\left(\frac{\Delta R}{0.3\,\rm km}\right)\left(\frac{\epsilon}{10^{-3}}\right)^2,
\label{eq:Estr}
\end{aligned}
\ee
where $\epsilon$ is the strain of the crust and $\mu\approx2.5\times10^{30}\,\rm erg\,cm^{-3}$ is the shear modulus of the crust \citep{1995MNRAS.275..255T}.
This energy release is consistent with the budget for an individual burst.

\begin{figure}
    \centering
    \includegraphics[width=0.96\linewidth]{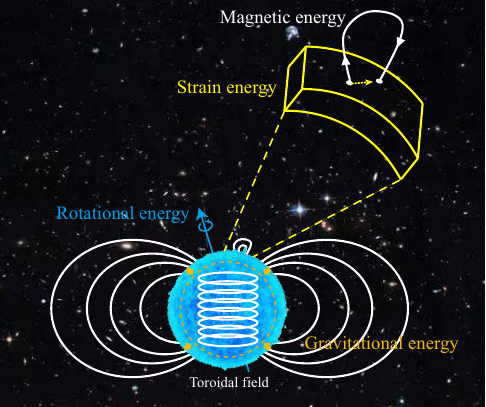}
    \caption{A schematic cartoon for energy stored and released in a magnetar. Starquakes can cause the footpoints of magnetic field lines to move, leading to magnetic reconnection. Gravitational energy may be released if the bulk shrinks, and part of it may convert into strain energy in the crust, and some strain energy may be released directly. Rotational energy can be released via quakes and spindown, with the latter being the direct energy source for FRBs.}
    \label{fig:magnetic}
\end{figure}

\begin{figure}
    \centering
    \includegraphics[width=0.96\linewidth]{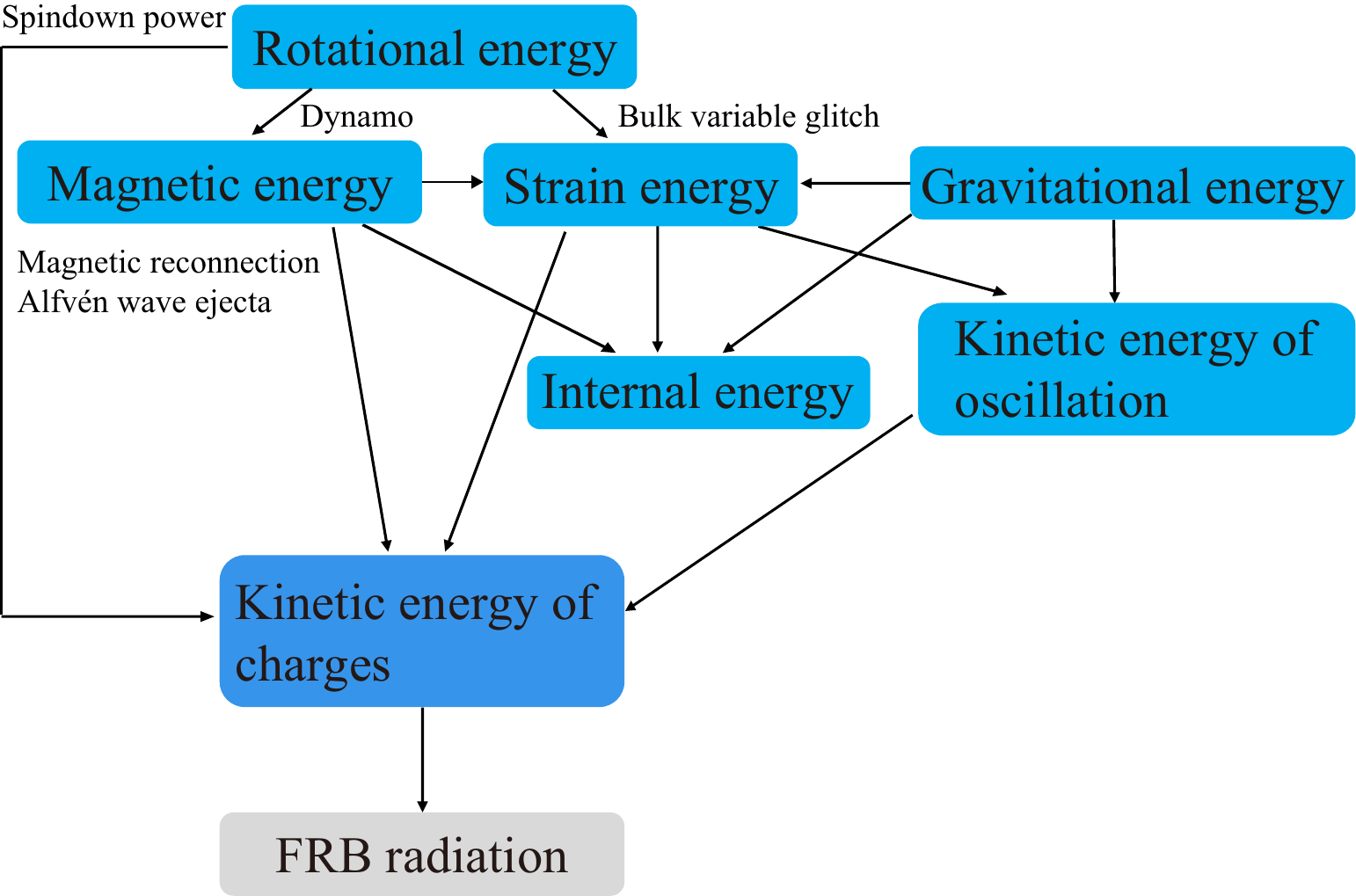}
    \caption{The energy evolutionary path for magnetic, rotational strain, and gravitational energy.}
    \label{fig:energy}
\end{figure}

During the crustal motion, some strain energy is accumulated and it would be released when the stress
exceeds a threshold.
The starquakes may be induced via some magnetic activities or crust collapse.
Also, starquakes may be associated with spindown glitches in which strain energy is accumulated via spindown-induced shape change.
These processes are also accompanied by magnetic, rotational, and gravitational energy release, which are summarized in Figure \ref{fig:magnetic}.
The possible processes involve a magnetar directly releasing these energies, partially converting these energies to strain energy.
Crustal motion can generate seismic waves leading to oscillations of the star.
The toroidal oscillation of the star enables the Goldreich-Julian density \citep{1969ApJ...157..869G} to deviate from the traditional value due to rotation, which provides additional voltage to accelerate charged particles to emit photons \citep{2015ApJ...799..152L}.
The energy evolutionary paths are summarized in Figure \ref{fig:energy}. Some situations occur alternatively and some processes take place successively or simultaneously.
We discuss the release mechanisms of these energies independently in the following.

\subsection{Magnetic energy released of starquake}\label{sec3.2}

Within the magnetar scenario, the magnetic energy can be released from the crust (and possibly even from the interior of the crust) into the magnetosphere, when the pressure induced by the internal magnetic field exceeds a threshold stress.
The magnetic energy dominates the crustal deformation energy if the magnetic field strength $B>B_\mu=(4\pi\mu)^{1/2}=5.6\times10^{15}\,\rm G$.
Suppose the magnetic field is deformed away from magnetostatic equilibrium by an amount of $\delta B$. In that case, the elastic stress suddenly balances the Maxwell stress, i.e., $\mu\epsilon\simeq B\delta B/(4\pi)$.
The largest yield strain of the crust is estimated as $\sim10^{-4}-10^{-2}$.
The effect of ohmic decay is neglected for young magnetars.

During the sudden crustal motion and starquake, an Alfv\'en wave packet is launched from the surface, and the wave vector is not exactly parallel to the field line and there is a non-zero electric current along the magnetic field lines \citep{2020MNRAS.494.2385K}.
The wave can become quite nonlinear and get ejected from the magnetosphere, if the wave packet propagates to a height in which total energy is greater than the magnetospheric energy \citep{2022ApJ...933..174Y,2023MNRAS.524.6024S}.
The ejecta pushes open the magnetic field lines and forms an electric current sheet connected back to the closed zone.
Magnetic reconnection can occur at the current sheets and then induce some X-ray burst counterparts.
The magnetic energy release is
\be
\begin{aligned}
\delta E_{\rm mag,c}&=1.2\times10^{40}\,{\rm erg}\left(\frac{\epsilon}{10^{-3}}\right)^2\left(\frac{B_s}{10^{15}\,\rm G}\right)^2\\
&\times\left(\frac{l}{10^4\,\rm cm}\right)^{2}\left(\frac{\Delta R}{0.3 \,\rm km}\right),
\end{aligned}
\label{eq:E_mag}
\ee
where $B_s$ is the magnetic field strength on the stellar surface and $\Delta R$ is the crustal thickness.
There is a height where charge density is insufficient to supply the current required by the wave and an electric field is formed to accelerate particle bunches, converting the magnetic energy into emissions.

The timescale of the energy release process in an area of $l^2$ is related to some instability growth.
A displacement of the magnetic footpoints on the surface can be caused by the diffusion of the internal magnetic field, which is reminiscent of the turbulent convective motions in the Sun.
The exchange in the positions of the footpoints is driven by the interchange instability \citep{1993ApJ...408..194T}.
We take the density of the neutron drip $\rho_n$ as the crustal density and then the Alfv\'en speed in the crust is written as $B_s/(4\pi\rho_n)^{1/2}$.
The growth time of the instability driven by the external magnetic field is 
\be
t_{\rm ex}\sim l/v_A\lesssim 2.2\times10^{-5}\,{\rm s}\left(\frac{l}{10^4\,\rm cm}\right)\left(\frac{B_s}{10^{15}\,\rm G}\right)^{-1}.
\ee
In the case where instability is driven by the internal field, the growth time is
\be
t_{\rm in}\sim 1.1\,{\rm ms}\left(\frac{l}{10^4\,\rm cm}\right)\left(\frac{B}{10^{15}\,\rm G}\right)^{-1}\left(\frac{\rho_c}{10^{15}\,\rm g\,cm^{-3}}\right)^{1/2},
\ee
where $\rho_c$ is the mass density of the interior region.

The solid crust of a magnetar plays a crucial role in various high-energy activities such as FRB.
Elastic stress accumulates with the evolution of the internal $B$-field and stores a large amount of energy in the form of elastic stress.
The total magnetic energy stored in the crust is
\be
E_{\rm mag, c}=1.5\times10^{46}\,{\rm erg}\left(\frac{B_s}{10^{15}\,\rm G}\right)^2\left(\frac{R}{10\,\rm km}\right)^2\left(\frac{\Delta R}{0.3 \,\rm km}\right),
\label{eq:E_mag,c}
\ee
where $R$ is the stellar radius.
Starquakes can occur anywhere on the star's surface in principle so that the global beam factor is close to an order of 1 even if it can be small.
The burst number supported by the crustal magnetic energy is estimated by
\be
\begin{aligned}
N&=1.0\times10^4\left(\frac{B_s}{10^{15}\,\rm G}\right)^2\left(\frac{R}{10\,\rm km}\right)^2\left(\frac{\Delta R}{0.3 \,\rm km}\right)\left(\frac{\eta}{10^{-4}}\right)\\
&\times\left(\frac{F_b}{0.5}\right)^{-1}\left(\frac{D}{1\,\rm Gpc}\right)^{-2}\left(\frac{\bar{F}_\nu}{0.3\,\mathrm{Jy}\,\mathrm{ms}}\right)^{-1}\left(\frac{\Delta \nu}{1\,\mathrm{GHz}}\right)^{-1}.
\end{aligned}
\label{eq:Nmag}
\ee
Considering possible duty cycles, the magnetic energy in the crust may not be sufficient to support the observed FRBs.
We take an average event rate of $100\,\rm h^{-1}$.
An FRB is assumed to be triggered at only one position so that the event rate is independent of the global beaming factor.
The magnetic energy in the crust can support consecutive FRB activities for $\approx100$ hours.

Consider a magnetic field with a general poloidal form of $B=B_sR^n/r^n$ ($n\geq3$).
The total magnetic energy stored in the magnetosphere is given by
\be
E_{\rm mag,m}=\frac{B_s^2R^3}{2(2n-3)}.
\ee
For a purely dipolar configuration ($n=3$), the total magnetic energy is 
\be
E_{\rm mag, m}=1.7\times10^{47}\,{\rm erg}\left(\frac{B_s}{10^{15}\,\rm G}\right)^2\left(\frac{R}{10\,\rm km}\right)^3.
\ee
The magnetic energy stored in the magnetosphere is higher than that stored in the crust and can support $\sim10^5$ bursts.
However, it is difficult to release such magnetic energy in a dipolar configuration, because the dipole is the ground state of the magnetic field.
The free energy that magnetic fields can release favors multipoles or twisted magnetic fields.

Some toroidal magnetic field components might anchor in a highly conducting crust of a neutron star \citep{2002ApJ...574..332T}.
The toroidal magnetic field can be transferred from the interior to the crust via crustal differential rotation along gravitational equipotential surfaces.
A sudden crust quake can twist the magnetic field in the outer magnetosphere \citep{2009ApJ...703.1044B}.
The magnetic energy of the twisted field is then launched into the magnetosphere.

Considering that the toroidal magnetic field in the crust is $10^{16}$ G, the energy can support $\sim10^6$ bursts based on Equation (\ref{eq:Nmag}).
The elastic strain can hardly balance the magnetic strain because of $B>B_{\mu}$ so that the crust is unstable and frequent starquakes may occur at the surface.
In this case, the magnetic energy stored in the crust is at least $6.3 \times10^{46}$ erg by taking $B=B_\mu$ in Equation (\ref{eq:E_mag}).
A simulation has found that if the magnetic field has a strong toroidal component, the crust is most prone to rupture, and in this case, the epicenter of crustal earthquakes is near the equator \citep{2015MNRAS.449.2047L}.
By assuming that the magnetic strain is comparable to gravity, there might be an upper limit of magnetic field of $\sim10^{18}$ G, leading to $\sim10^{10}$ bursts, which is a strict upper limit.

\subsection{Rotational energy release via quake and spindown}\label{sec3.3}
We discuss the release of rotational energy via a global glitch and directly convert it into radiations.

\subsubsection{Spindown-induced glitch with starquake}\label{sec3.3.1}
A global starquake may happen when the shear strain due to a spindown-induced shape change exceeds a threshold.
Note that a neutron star is thought to be a fluid star with a thin solid crust.
The equilibrium configuration of a rotating neutron star leads to an ellipsoid rather than a perfect sphere, in which elastic energy is accumulated in the crust \citep{1971AnPhy..66..816B}.
This global starquake is also associated with a glitch and consists of three steps: (1) the normal spindown phase, which begins at the end of the last glitch and during which the elastic energy is accumulated; (2) the glitch epoch, during which the star loses its elastic energy; and (3) the glitch phase, during which the star changes its shape and sets up a new equilibrium at the end of this phase \citep{2004APh....22...73Z,2008MNRAS.384.1034P,2014MNRAS.443.2705Z}.
The total energy of the star is given by \citep{1971AnPhy..66..816B}
\be
E_{\text {total }}=E_{\mathrm{k}}+E_{\mathrm{g}}+E_{\text {ela }}=E_0+\frac{I\Omega^2}{2}+A \varepsilon^2+B_v\left(\varepsilon-\varepsilon_0\right)^2,
\label{eq:Eglitch}
\ee
where $I$ is the neutron star's moment of inertia, $\Omega$ is the angular velocity, and
\be
\begin{aligned}
&A=\frac{3 G M^2}{25 R},\\
&B_v=\frac{2\pi R^3\mu}{3},\\
&\varepsilon=\frac{I-I_0}{I_0}.
\end{aligned}
\ee
The gravitational and kinetic energy changes correspond to the shape change and spindown.
However, the ellipticity change during the glitch is small, and the total energy for the glitch epoch is approximated as that of the glitch phase.
The total energy change during the quake is mainly the strain energy, which is calculated as \citep{2014MNRAS.443.2705Z}
\begin{equation}
E_{\rm str,g}=\frac{A}{2(A+B_v)} |\delta E_k|\frac{\delta \Omega}{\Omega}.
\end{equation}
The upper limit for the total strain energy is the total spin energy.

After the previous quake, the stress in the crust can build up again.
The rate of stress build-up is
\be
\dot{\sigma}=-\mu \dot{\varepsilon}=-\frac{8\pi^2\mu I_0\dot{P}}{P^3(A+B_v)},
\ee
where $P$ is the rotation period.
Therefore, the waiting time for the starquakes can be estimated as $t_q=\sigma_c/\dot{\sigma}$.
We take $\delta\Omega/\Omega=10^{-9}$, leading to $\sigma_c\approx\mu\delta\Omega/\Omega=2.5\times10^{21}\,\rm erg\,cm^{-3}$.
Considering $A\gg B_v$, we can obtain
\be
t_q=19.7\,{\rm h}\left(\frac{\delta\Omega/\Omega}{10^{-9}}\right)\left(\frac{P}{1\,\rm s}\right)^4\left(\frac{B_s}{10^{15}\,\rm G}\right)^{-2}\left(\frac{R}{10\,\rm km}\right)^{-7}.
\ee
The waiting time for the starquakes is much longer than the waiting time for some short-interval FRBs of $\sim10\,\rm ms-100\,s$.
Strain energy is therefore not a viable FRB power source since it cannot be released frequently.

\subsubsection{Rotational energy for young mangetars}\label{sec3.3.2}
Regardless of whether the radiation mechanisms of pulsars and FRBs are the same, we consider the spindown power of a neutron star to sustain FRBs.
The spindown power is independent of a starquake scenario and is just regarded as an energy mechanism in the following discussion.

The rotational energy of a neutron star can sustain radio emission from a pulsar \citep{1968Natur.218..731G}.
The strong magnetic field and high rotation speed of neutron stars cause the magnetosphere to fill with plasma, leading to radiation in the pattern of a rotating beacon \citep{1969ApJ...157..869G}.
The total spin energy of a rotating neutron star with $M=1.4M_\odot$ is given by
\be
E_{\rm rot}=\frac{1}{2}I\Omega^2
=2.2\times10^{46}\,{\rm erg}\left(\frac{R}{10\,\rm km}\right)^2\left(\frac{P}{1\,\rm s}\right)^{-2}.
\ee
The neutron star can be regarded as a rotating magnetic dipole.
%in vacuum.
The corresponding radiation luminosity is defined by its spindown luminosity, which is estimated by
\begin{equation}
\begin{aligned}
\dot{E}_{\mathrm{d}}& \simeq \frac{2 B_s^2 R^6\Omega^4}{3 c^3} \\
%\sin ^2 \alpha\\
&=3.8\times10^{37}\,{\rm erg\,s^{-1}}\left(\frac{B_s}{10^{15}\,\rm G}\right)^2\left(\frac{R}{10\,\rm km}\right)^6\left(\frac{P}{1\,\rm s}\right)^{-4},
\end{aligned}
\end{equation}
where we do not consider an inclination angle dependence because the power is insensitive to the angle when both magnetic dipolar spindown and wind spindown are considered \citep{1999ApJ...525L.125H,2001ApJ...561L..85X,2006ApJ...648L..51S}. The corresponding radiation luminosity should be equal to the spindown power, which gives the period evolution of the neutron star:
\begin{equation}
\dot{P}=\frac{8 \pi^2 B_s^2R^6}{3 I c^3} P^{-1}.
\end{equation}
The spindown age is 
\be
t_{\rm sd}=\tau_{\mathrm{c}}\left[1-\left(\frac{P_0}{P}\right)^2\right],
\ee
where $P_0$ is the initial rotational period.
It can be reduced to $\approx P/(2\dot{P})$ when the initial rotational period $P_0$ is much smaller than $P$.

For a typical magnetar with $P\sim1$ s, the rotational energy is in the same order as the total magnetic energy stored in the crust, given as Equation (\ref{eq:E_mag,c}).
According to discussions in Section \ref{sec3.3}, the rotational energy can sustain $\sim10^4$ bursts.
However, the spindown power of a magnetar with $P\sim1$ s may not be able to support very bright bursts whose observed isotropic luminosities greatly exceed the spindown luminosity.

We consider a magnetar with $P=0.1$ s, which is the point of interest in the following discussion.
The spindown power is sufficient for the FRB luminosity.
However, the spindown age is calculated to be $\sim$ 0.2 years which is much smaller than the age since the FRBs were discovered.
There might be a lot of bursts missing before the FRB source was first discovered.
As the rapid spindown evolution, the luminosity of FRB decreases significantly.

Such a newly born magnetar is surrounded by a supernova remnant.
We take the spindown age as the age of supernova remnant.
The free-free optical depth for the wave travelling through the ejecta shell ($Z = 8$, $A = 16$) for an oxygen-dominated composition) is given by \citep{2017ApJ...841...14M}
\be
\begin{aligned}
\tau_{\rm ff}&=0.018Z^2\nu^{-2}T_{\rm ej}^{-3/2}n_en_ig_{\rm ff}R_{\rm ej}\\
&=9.2\times10^{10}g_{\rm ff}\left(\frac{\nu}{1\,\rm GHz}\right)^{-2}\left(\frac{T_{\rm ej}}{10^4\,\rm K}\right)^{-3/2}\left(\frac{M_{\rm ej}}{10\,M_\odot}\right)^2\\
&\times\left(\frac{v_{\rm ej}}{10^4\,\rm km\,s^{-1}}\right)^{-10}\left(\frac{P}{0.1\,\rm s}\right)^{-10}\left(\frac{B_{s}}{10^{15}\,\rm G}\right)^{10},
\label{eq:tau}
\end{aligned}
\ee
where $g_{\rm ff}\simeq1$ is the Gaunt factor, $T_{\rm ej}$, $M_{\rm ej}$ and $v_{\rm ej}$ are the temperature, mass, and mean velocity of the ejecta.
For low-frequency photons, the induced Compton scattering is significant if the flux is high enough \citep{1978MNRAS.185..297W,2008ApJ...682.1443L,2020ApJ...893L..26I}, which gives the optical depth:
\be
\begin{aligned}
\tau_{\rm C}&\simeq \frac{3 \sigma_T}{32 \pi^2} \frac{n_e \delta E}{R_{\rm ej}^2 m_e \nu^3}\\
&=8.0f_b \left(\frac{D}{1\,\rm Gpc}\right)^2\left(\frac{F_\nu}{0.1\,\mathrm{Jy}\,\mathrm{ms}}\right) \left(\frac{\Delta \nu}{1\,\mathrm{GHz}}\right)\\
&\times\left(\frac{M_{\rm ej}}{10\,M_\odot}\right)\left(\frac{\nu}{1\,\mathrm{GHz}}\right)^{-3}\left(\frac{v_{\rm ej}}{10^4\,\rm km\,s^{-1}}\right)^{-5}\\
&\times\left(\frac{R}{10\,\mathrm{km}}\right)^{20}\left(\frac{B_s}{10^{15}\,\mathrm{G}}\right)^{10}\left(\frac{P}{0.1\,\mathrm{s}}\right)^{-10},
\end{aligned}
\ee
where $\sigma_T$ is the Thomson scattering cross-section and $m_e$ is the mass of an electron.
GHz waves from a magnetar with $B_s=10^{15}$ G and $P=0.1$ s are optically thick and cannot escape from the ejecta shell.

\begin{figure}
    \centering
    \includegraphics[width=0.96\linewidth]{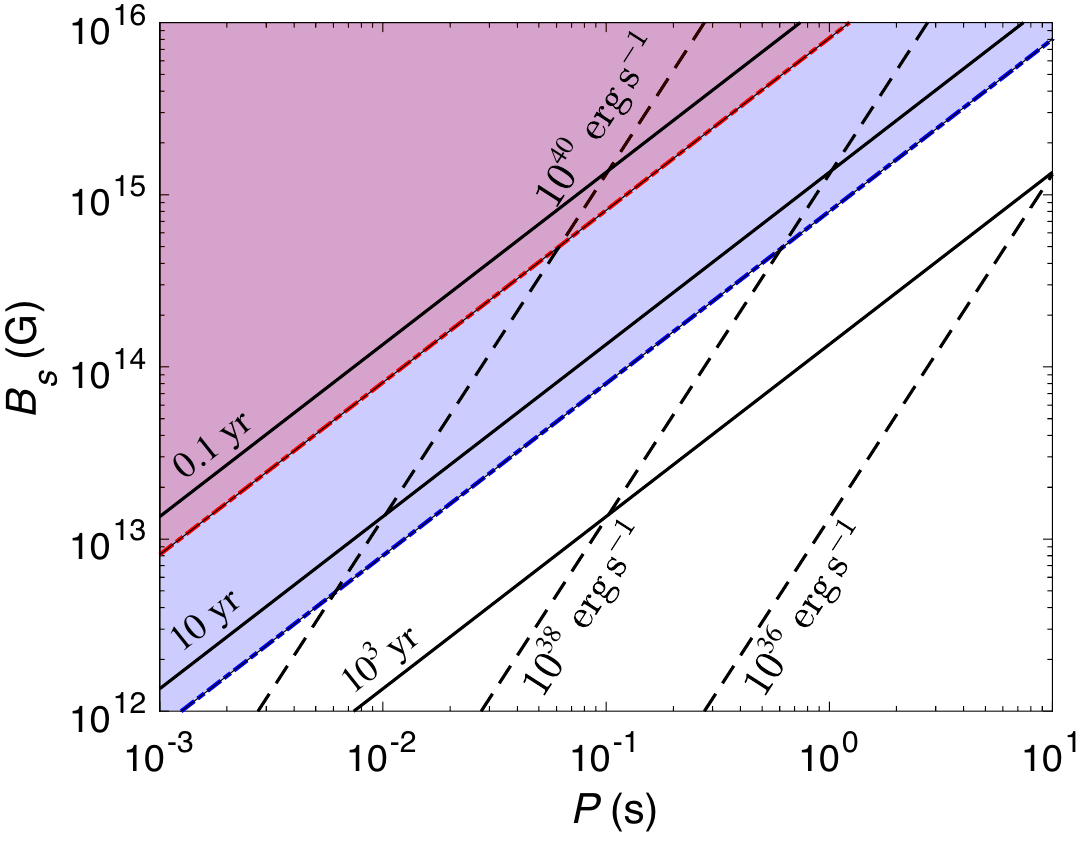}
    \caption{$P-B_s$ diagram. We take $M_{\rm ej}=10M_\odot$, $v_{\rm ej}=10^4\,\rm km\,s^{-1}$, $\nu=10^9$ GHz, $T_{\rm ej}=10^4$ K. The black dashed lines denote spindown energy rates which $\dot{E}_d=10^{36},10^{38},10^{40}\,\rm erg\,s^{-1}$. The black solid lines denote spindown ages which are 0.1, 10, and $10^3 $ yr. The red and blue dashed-dotted lines denote $\tau_C=1$ and $\tau_{\rm ff}=1$. The red and blue regions denote the optically thick for induced Compton scattering and free-free absorption.}
    \label{fig:tau}
\end{figure}

We plot the relationship between $P$ and $B_s$, as shown in Figure \ref{fig:tau}.
Both the parameter spaces of optically thick are shown as the color regions for the free-free absorption and induced Compton scattering.
A magnetar with $P\gtrsim0.1$ s can sustain GHz FRBs and the bursts can escape from the ejecta shell region.

The ejecta shell is too dense to let FRB waves escape a newly-born magnetar.
Hence, a magnetar formed by a compact binary merger may host an optically thin shell for FRBs which has a lower ejecta mass.
The magnetar formation matches the case of FRB 20200120E localized to an old globular cluster in M81 \citep{2021ApJ...910L..18B,2021ApJ...917L..11K,2021ApJ...919L...6M,2021ApJ...917L..11K}.
Nevertheless, whether a compact binary can form a magnetar lacks observational evidence.
Massive star core collapses are likely the main pathway for the formation of the majority of young magnetars.

\subsection{Gravitational and phase transition energies}\label{sec3.4}
We consider gravitational energy released during starquakes.
A hadron-to-quark phase transition can convert a neutron star to a quark star or a hybrid star in realistic scenarios like spindown, depending on whether the quark matter phase is absolutely stable or not. Such a phase transition is accompanied by a release of energy from the changes of internal energy and gravitational potential energy of roughly the same order of magnitude~\citep{bombaci2000conversion}, while the release of gravitational potential energy can be approximately evaluated as
\be
\begin{aligned}
E_{\rm gra}&=\frac{3GM^2\delta R}{5R^2}\\
&=3.3\times10^{47}\,{\rm erg}\left(\frac{M}{1.4\,\mathrm{M}_{\odot}}\right)^2\left(\frac{R}{10\,\mathrm{km}}\right)^{-1}\left(\frac{\delta R/R}{10^{-6}}\right),
\end{aligned}
\label{eq:E_gra}
\ee
where $\delta R$ is a global reduction in radius.
When converting to hybrid stars, the phase transition and the associated energy release have been shown to be able to source core-quakes~\citep{Bejger:2005am} and repeated fast radio bursts~\citep{Shen:2023wid}. Recently, it was shown that a phase transition from strangeon matter \footnote{The building blocks of a strangeon star are strangeons which are localized quark clusters formed by the strong force, in analogy of atomic nucleons but with a large baryon number and with strange quarks in almost equal fraction as up and down quarks.} to strange quark matter is feasible in the core region of strangeon stars~\citep{Zhang:2023szb}, which may also result in core quakes and release of a large amount of energy due to the gravitational mass changes.

Starquakes can induce pressure anisotropy changes inside a star, which in turn can cause a large amount of energy released from associated gravitational mass changes~\citep{xu2006superflares,chen2024free} even without phase transition. It has been shown that a small amount of anisotropy change of the order of $10^{-4}$ will release a huge amount of energy ($>10^{46}$ erg)~\citep{chen2024free}.
%possibly in a way of coherent emission from magnetic disturbances induced by the phase transition, as described in~\citep{2020MNRAS.498.1397L}.
% the latter of which demonstrated how a repeated process of phase transitions can lead to repeated FRBs and glitches that are consistent with observations.
However, pressure resists gravity and does work during the collapse, converting most of all gravitational energy into internal energy.
The efficiency of gravitational energy conversion into radiation seems to be small, unless the shear modulus of the star is large.

The radiation mechanism from gravitational energy to radiation is not very determined.
One possible mechanism is that the star has oscillations induced by seismic waves.
The oscillations can enhance the voltage potential on the polar gap causing more energetic emissions \citep{2015ApJ...799..152L}.
A part of the released gravitational energy is transformed into kinetic energy of the oscillations and then becomes radiation power.
However, it is not clear how much energy is converted into thermal energy during core heating via quakes.
The gravitational energy budget seems to give a theoretical upper limit on the energy released from a pulsar-like compact star.

Another interesting scenario is relevant to the challenging equation of state of cold supra-nuclear matter, in which huge gravitational energy may efficiently power emission if the equation of state is so stiff that the maximum mass, $M_{\rm TOV}$, calculated with the Tolman-Oppenheimer-Volkoff equation reaches $(2.5\sim 3.0)M_\odot$, particularly after binary neutron star merger.
The mass of a merger remnant could even be slightly higher than $M_{\rm TOV}$ for two companion stars with typical masses of $\sim 1.5M_\odot$, but the hyper-massive neutron star would survive due to rapid rotations before collapsing into a black hole.
Besides the centrifugal force, in a strangeon star model, an elasticity would additionally prevent the remnant from collapsing after solidification~\citep{2017RAA....17...92Y}, and quakes could then occur frequently afterward.
This kind of massive remnants could also be attractive gravitational wave sources, even echoes~\citep{2023PhRvD.108f3002Z}, with multi-messengers.
In any case, there are quantitative uncertainties in this scenario to be investigated in the future.

Alternatively, another way to release gravitational energy is for neutron stars to accrete external matter, e.g., asteroid \citep{2016ApJ...829...27D,2018ApJ...858...88Z,2021Innov...200152G}.
The accreted matter falls onto the stellar surface leading to starquakes.
During this process, the gravitational energy is firstly converted into kinetic energy and then transformed into emitting charges via somehow mechanisms.
However, it is hard to interpret how frequently the bursts can happen, e.g., the event rate can be up to $100\,\rm h^{-1}$.

\section{Discussion}
\label{sec4}

\subsection{Narrow spectra and impact on energy distribution}
\label{sec4.1}
The bandwidth limitations may affect the calculation of burst energy.
Observations of repeaters show that the energy/luminosity function spans in a wide range.
The distribution has a high energy cut-off, while no sharp cut-off at the low energy bands due to sensitivity limitations.
For some radio telescopes, the limited sensitivity only allows to catch a small number of bright bursts whose function can be well modelled by a power law distribution.
If the sensitivity is high enough, one can see hidden features.
%FAST's observations indicate that the energy functions from different sources vary, suggesting a characteristic unique to each source.

For some FRBs, radiation frequencies appear to exhibit distinct preferences, being particularly active at certain frequencies.
This phenomenon may lead to a significant deviation between the distribution observed by narrowband telescopes and the intrinsic distribution.
Since repeaters have a relatively small bandwidth ($\sim200$ MHz (e.g. FRB 20220912A, see \citealt{2023ApJ...955..142Z}), if the central frequency of a burst is outside of the observing band of the telescope, only a small portion of the emission energy is detected. 
According to Equation (\ref{eq:E_obs}), this gives an under-estimated burst energy. Such an issue brings challenges to obtain robust energy estimation and introduces a bias to the observed energy distribution.

\subsection{Duty Cycle}
\label{sec4.2}
We simply discuss the duty cycle by considering constant burst rate during the lifetime.
However, for most repeaters, we are not sure whether they are physically inactive or due to the observational sensitivity, and the constant burst rate in terms of time is a strong hypothesis.
Some repeaters may have periodically active windows (e.g., \citealt{2020Natur.582..351C,2020MNRAS.495.3551R}).
If the true duty cycle is small and a high value is adopted as the average, we may overestimate the energy budget.
A detailed burst rate evolution in time needs to be considered.

\subsection{Period for energetic FRBs?}
\label{sec4.3}
Many FRB models invoke rotating compact stars. However, the spin period has not been found from burst timing analyses (e.g., \citealt{2022RAA....22l4004N}).
FRBs are much more energetic than radio pulses from normal pulsars, so the well-defined open field line region is likely distorted and enlarged.
Within the framework of a magnetar, a starquake can lead to twisting of the magnetic field, and the twisted current-carrying bundle plays a role similar to the open field line region of normal pulsars.
The front of fireballs due to the X-ray bursts can push aside the pair plasma of closed field lines, forming a clean region to let radio waves escape \citep{2020ApJ...904L..15I}.
The observed duty cycles are then enlarged and the emission phase is more random, matching the observation of FRB-like bursts from SGR J1935+2154 \citep{2023SciA....9F6198Z}.

The missing period of FRBs may have a strong relationship with the trigger mechanism.
We speculate that the charged emitting particles can be triggered by point discharges from ``hills'' on the stellar surface.
The hills may be created by crust shear and motion which may be randomly distributed on the surface.
Some hills are higher and sharper, allowing them to generate more charges, and emitting brighter pulses.
These higher hills may stand more stably during frequent crustal activities, unlike the shorter hills that may disappear.
As a result, some bright bursts can have stable phases, and timing measurements of them could display spin period modulation (e.g., \citealt{2024arXiv240416669L}).
As shown in Figure \ref{fig:tau}, if the magnetar was born from a core-collapse supernova,  the spin period cannot be much shorter than $\sim1$ s, unless the bursts are too faint.
The non-detection of periodicity from FRBs does not necessarily rule out a rotating compact star as the FRB source.

\section{Summary}
\label{sec5}
In this work, we discuss the energy budget of FRBs and investigate various energy sources and dissipation mechanisms related to starquakes.
During the starquake, the strain energy can be directly released and possibly accompanied by the other three kinds of energy releasing and conversion.
We have drawn the following conclusions:

1. An Alfv\'en wave packet can be launched from the surface during starquakes forming an electric field along the magnetic field lines, accelerating charged particles to a streaming outflow, so that magnetic energy is converted to emission.
The magnetic energy released within a patch size of $l=10^4$ cm is sufficient for an FRB.
By considering a purely poloidal component of the magnetic field, the energy stored in the crust can support roughly $10^4$ bursts.
More free energy can be released if there are toroidal components inside the magnetar.
The crust might be unstable with frequent starquakes happening if $B>B_\mu$, suggesting a magnetic energy at least of $6.3\times10^{46}$ erg in the crust.

2. For a neutron star with a solid crust and a stiff equation of state, the strain energy can accumulate during the neutron star spindown and is suddenly released during a glitch.
However, the star needs to spend a long time to accumulate an enough strain energy to induce a starquake. This time is much longer than the observed waiting times of FRBs.

3. The total rotational energy of a normal magnetar with $P\sim1$ s is comparable to the magnetic energy stored in the crust, while the spindown power cannot support some bright FRBs.
The rotational energy of a magnetar whose period is comparable to 0.1 s or shorter can match the energy and luminosity budget of FRBs. However, the GHz radio emission cannot escape from the ejecta shell region for an extremely young magnetar due to the high density. The spindown timescale is also too short compared with the observed lifetimes of active repeaters. 

4. For the neutron star scenario, quakes of a quark matter core from phase transitions can release gravitational energy and phase transition energy.
For the solid quark star (strangeon star) scenario, starquakes can happen frequently and may release a huge amount of gravitational energy if the equation of state is so stiff that the maximum mass can be as high as $(2.5\sim 3.0)M_\odot$, allowing a stable merger product from a binary merger.
The total gravitational energy is high but it is not known whether such energy can be converted to emissions with a large enough efficiency. 
This energy budget seems to give a theoretical upper limit of energy release of a pulsar-like compact star.

\normalem
\begin{acknowledgements}
The authors thank the anonymous referee for valuable suggestions.
We are grateful to Shunshun Cao, Mingyu Ge, Kejia Lee, Bing Li, Ze-Nan Liu, Yudong Luo, Chenhui Niu, Yuan-Pei Yang, Xiaoping Zheng, Xia Zhou, and members of ``DJ Find Girlfriend Assistant'' Wechat group for helpful discussion.
This work is supported by the National SKA Program of China (No. 2020SKA0120100) and the Strategic Priority Research Program of the CAS (No. XDB0550300).
J.F.L. acknowledges support from the NSFC (Nos.11988101 and 11933004) and from the New Cornerstone Science Foundation through the New Cornerstone Investigator Program and the XPLORER PRIZE.

\end{acknowledgements}

\bibliographystyle{raa}
\bibliography{bibtex}
% \begin{thebibliography}{39}
% \providecommand\natexlab[1]{#1}
% \providecommand\JournalTitle[1]{#1}
% \end{thebibliography}

%\appendix
\include{table}

\end{document}